\documentclass{aastex63}

\received{} 
\revised{} 
\accepted{} 
\submitjournal{ApJ}

\shorttitle{Waves and Flows in a Prominence Foot}
\shortauthors{Ofman and Kucera}

\begin{document}

\title{Fast Magnetosonic Waves and Flows in a Solar Prominence Foot: Observations and Modeling}

\correspondingauthor{Leon Ofman}
\email{ofman@cua.edu}

\author[0000-0003-0602-6693]{Leon Ofman}
\affiliation{Department of Physics\\
Catholic University of America  \\
Washington, DC 20064, USA}
\affiliation{NASA Goddard Space Flight Center \\
Code 671 \\
Greenbelt, MD 20771, USA}

\author[0000-0001-9632-447X]{Therese A. Kucera}
\affiliation{NASA Goddard Space Flight Center \\
Code 671 \\
Greenbelt, MD 20771, USA}




\begin{abstract} 
We study recent observations of propagating fluctuations in a prominence foot with Hinode Solar Optical Telescope (SOT) high-resolution observations in Ca~II and H$\alpha$ emission which we identify as nonlinear fast magnetosnic waves. Here we analyze further the observations of propagating waves and flows with Interface Region Imaging Spectrograph (IRIS)  Mg~II slit jaw images, in addition to Hinode/SOT Ca~II images. We find that the waves have typical periods in the range of $5 - 11$ minutes and wavelengths in the plane of the sky (POS) of about 2000~km, while the flows in narrow threads have typical speed in the POS of $\sim16-46$ km s$^{-1}$. We also detect apparent kink oscillations in the threads with flowing material, and apply coronal seismology to estimate the magnetic field strength in the range 5-17 G. Using 2.5D MHD we model the combined effects of nonlinear waves and flows on the observed dynamics of the prominence material, and reproduce the propagating and refracting fast magnetosonic waves, as well as standing kink-mode waves in flowing material along the magnetic field. The modeling results are in good qualitative agreements with the observations of the various waves and flows  in the prominence foot, further confirming coronal seismology analysis  and improving the understanding of the fine scale dynamics of the prominence material.

\end{abstract}



\section{Introduction} \label{intro:sec}

The magnetic structure of solar prominences is complex and not well understood. Evidence for flows, oscillations, and MHD instabilities  are seen in the cool prominence material \citep[see the review,][]{Par14}. High resolution observations of prominences with the Solar Optical Telescope (SOT) on board the Hinode satellite \citep{Kos07} reveals the complexity of the prominence dynamics \citep[e.g.,][]{Oka07,Ber08,Ber10,Hil13,Hin19}. Recent observations of the prominence structure and dynamics with the Solar Dynamics Observatory's (SDO) Atmospheric Imaging Assembly (AIA) \citep{Lem12} further revealed the complexity of the prominence megnetohydrodynamics  \citep[e.g.,][]{Liu12a,Ber12}. Observations with Interface Region Imaging Spectrograph (IRIS), with slit jaw images (SJI) and \ion{Mg}{2} spectral data \citep{DeP14} provide fine-scale  spectroscopic imaging and dynamics information of the cool prominence material \citep{Sch14,Hei15,Liu15,Via16,Oka16,Ant18,KOT18}.

Wave and oscillations in the magnetized prominence structures  have been observed for decades using ground-based and space-based telescopes. Due to the magnetized nature of the prominence material, the waves and oscillations were interpreted in terms of various MHD modes with broad range of scales  \citep[see the reviews by][]{Lin11,Arr18}. The various types of large- and small-scale oscillations and associated idealized linear MHD modes  allow the straightforward  application of coronal seismology in many instances \citep[see the review][]{NV05}. However, when  nonlinearity becomes important, for example, in large amplitude oscillations in prominences \citep[see the review,][]{Tri09}, nonlinear MHD modeling is needed to properly account for the nonlinear coupling and dynamics of these waves \citep[e.g.,][]{OKKS15}. The dissipation of MHD waves can also play an important role in the heating of the prominence material \citep[e.g.,][]{OM96,Ofm98b,Oka15}.

 \citet{OKKS15} studied the propagations of nonlinear waves in a prominence foot using a 2.5D MHD model and high-resolution Hinode/SOT observations in in Ca~II emission line of a prominence seen on 2012 October 10. They found highly dynamic small-scale propagating features in the prominence material, also evident in H$\alpha$ intensity and Doppler shifts. The feature propagating upward along the foot were interpreted as nonlinear fast magnetosonic waves with typical period in the range of 5 -- 11 minutes and wavelengths on the order of $\sim$2000 km or less. Previously, these waves were interpreted as linear fast magnetosonic waves \citep{Sch13}. The nonlinear 2.5D MHD model of propagating quasi-periodic nonlinear fast magnetosonic waves was able to capture the main features of observed propagating fluctuations, with the magnetic field intensity in the model guided by the THEMIS (T\'elescope H\'eliographique pour l'Etude du Magn\'etisme et des Instabilit\'es Solaires) \citep{Lop00} instrument observations. However, the above study did not consider the effects of flows that apparently cross the prominence foot, and the effects of gravity were only partially included. 

Mass flows of the cool prominence material are observed ubiquitously in prominence threads (see the review  by \citet{Par14}, and recent observations by \citet{Ale13,Kuc14,Die18}). Combined flows in small-scale threads and waves were also detected and studied in solar prominences by \citet{Oka07} and in coronal loops with cool prominence material by \citet{OW08}. Recently, \citet{KOT18} analyzed motions in prominence barbs observed on the solar limb with the Hinode/SOT in Ca~II, H$\alpha$,  IRIS SJI  Mg~II emission, and the SDO/AIA. They found line-of-sight fluctuations in the barbs and features moving along the barbs perpendicular to the solar limb that  tend to come in clusters along similar paths in the plane of the sky. Similar to previous observations \citep[e.g.,][]{Sch13,OKKS15}, they have detected the quasi-periodic wave-like features in prominence foot (or `pillar') that appear on timescales of 10 minutes to an hour, propagating toward or away from the limb with corresponding Doppler shift signatures.

In the next section we discuss the prominence observations. In Section~\ref{model:sec} we present the model and in Section~\ref{num:sec} the numerical results. Section~\ref{dc:sec} contains a summary and our conclusions.

\section{Observational Data Analysis} \label{obs:sec}

The observational data set and its analysis method were described in detail in \cite{KOT18}. To summarize, the prominence was observed on the west limb south of the equator on 2016 January 7 by the Hinode/SOT \citep{Kos07,Tsu08} in both \ion{Ca}{2} and H$\alpha$ and also by the IRIS, with SJI and \ion{Mg}{2} spectral data \citep{DeP14}.  In this paper we show examples of the SOT \ion{Ca}{2} 3969~\AA\  images and the IRIS \ion{Mg}{2}~k 2796~\AA\ SJI of the prominence pillar region used for the flows and waves analysis. 

The SOT \ion{Ca}{2} images were taken with a cadence of 30 s over a field of view of $76\times76\arcsec$ with a spatial resolution (2 pixels) of 0.21\arcsec. The data were taken over three time intervals, 07:17 - 07:34 UT, 08:55 - 09:13 UT, and 10:23 - 11:00 UT,  chosen to avoid the high particle count periods of the South Atlantic Anomaly (SAA).
The IRIS \ion{Mg}{2}k slit-jaw images have a 37.6~s cadence, a field of view of $168\times177$\arcsec, and a resolution (2 pixels) of 0.33\arcsec. They were taken over the intervals 07:29 - 08:24, 08:56 - 10:00, and 10:33 - 11:23~UT.
\begin{figure}[ht]
\centerline{
\includegraphics[width=\linewidth]{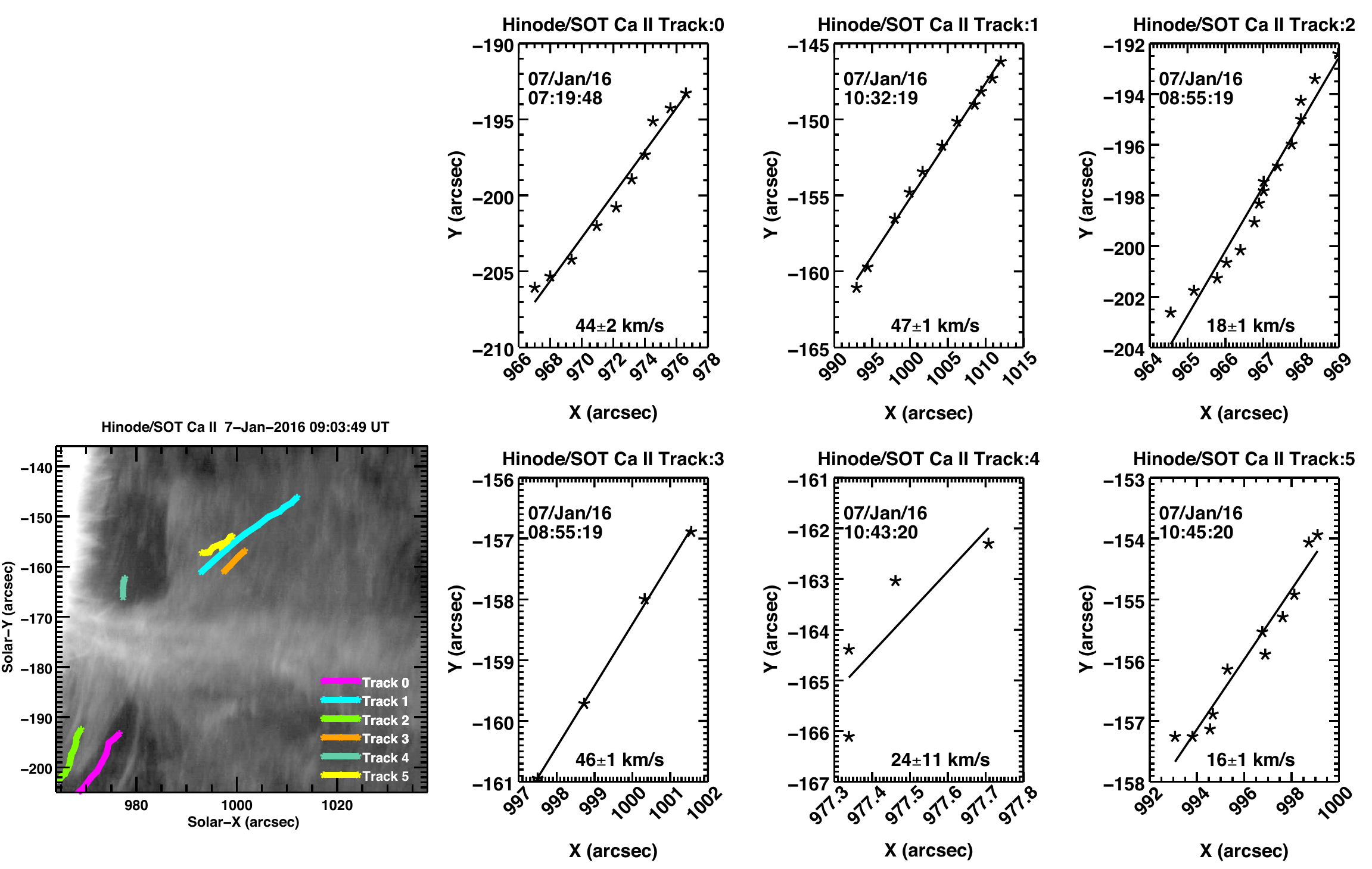}}
\caption{Left: Hinode/SOT Ca~II image of the prominence on 7-Jan-2016 09:03:49 UT with the colored tracks showing  the locations of the flow velocity measurements. (An animation of this panel is available online). Right: the velocities with error bars of the prominence material measured along the tracks. The time given is the time of the first frame used to track the feature.}
\label{CaIItracks:fig}
\end{figure}

\begin{figure}[ht!]
\centerline{
\includegraphics[angle=90,width=\linewidth]{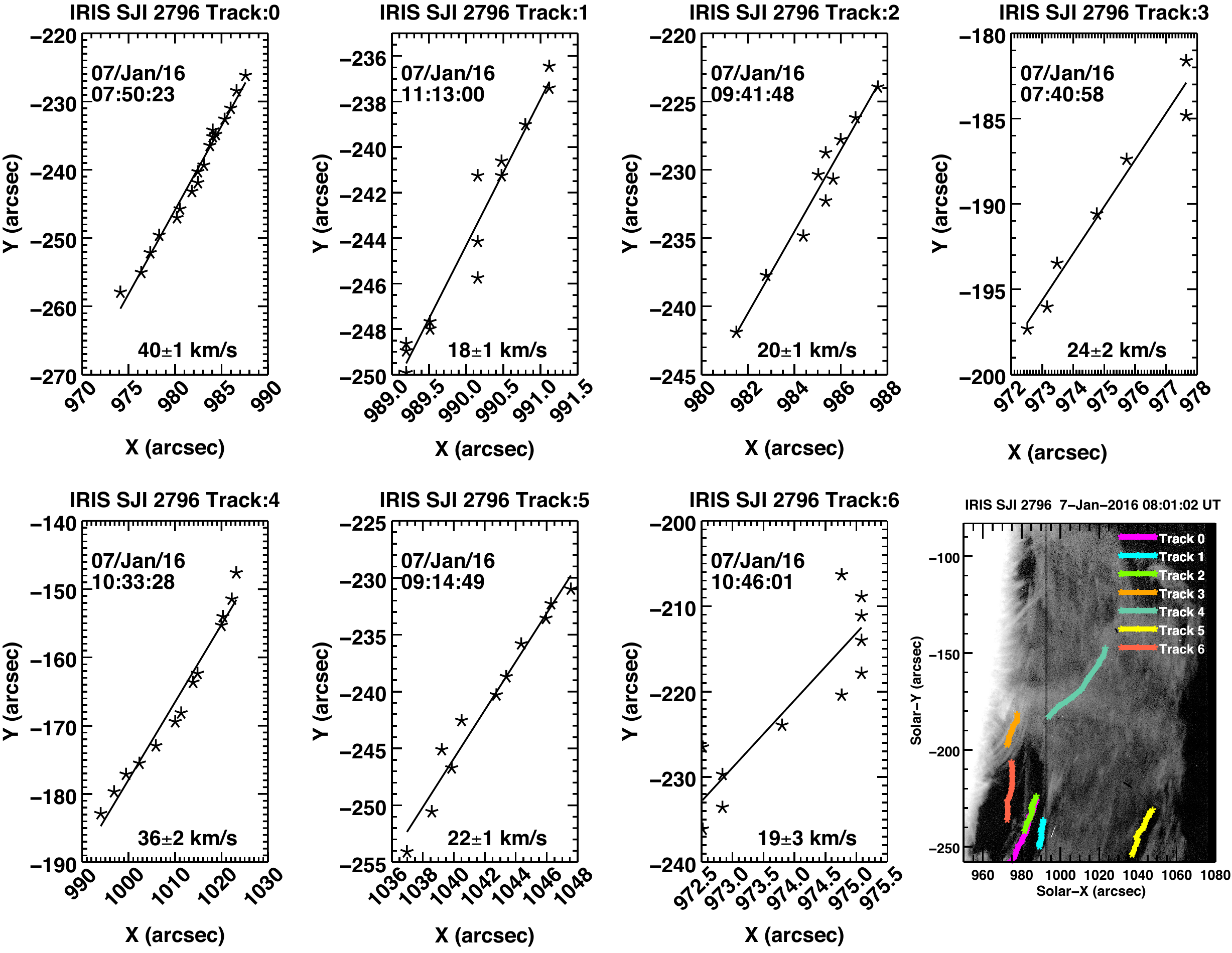}}
\caption{IRIS SJI in Mg~II emission of the prominence on 7-Jan-2016 08:01:02 UT with the colored tracks showing the locations of the flow velocity measurements (lower right panel)  (An animation of this panel is available online). The velocities with error bars of the prominence material measured along the tracks in IRIS Mg~II SJI. The time given is the time of the first frame used to track the feature.}
\label{MgIIv:fig}
\end{figure}

For this paper we obtained the velocities of the prominence material in the plane of the sky by measuring the motions of features that could be detected either in the \ion{Ca}{2} or \ion{Mg}{2}k in two ways. For one set of features in each waveband we tracked individual moving features manually. The locations of the features representing the flows, best seen outside the pillar structure are shown in  Figure~\ref{CaIItracks:fig} for \ion{Ca}{2} emission and Figure~\ref{MgIIv:fig} for \ion{Mg}{2} emission. We then preformed linear least-square fits to the points  with the resulting line slopes providing speeds in the plane of the sky from 16 to 47 km s$^{-1}$. The plane of the sky velocities provide lower limits of the true velocities due to the possible effects of the projection angle on the inclined propagation of the prominence material. 

\begin{figure}[ht!]
\centerline{
\includegraphics[width=\linewidth]{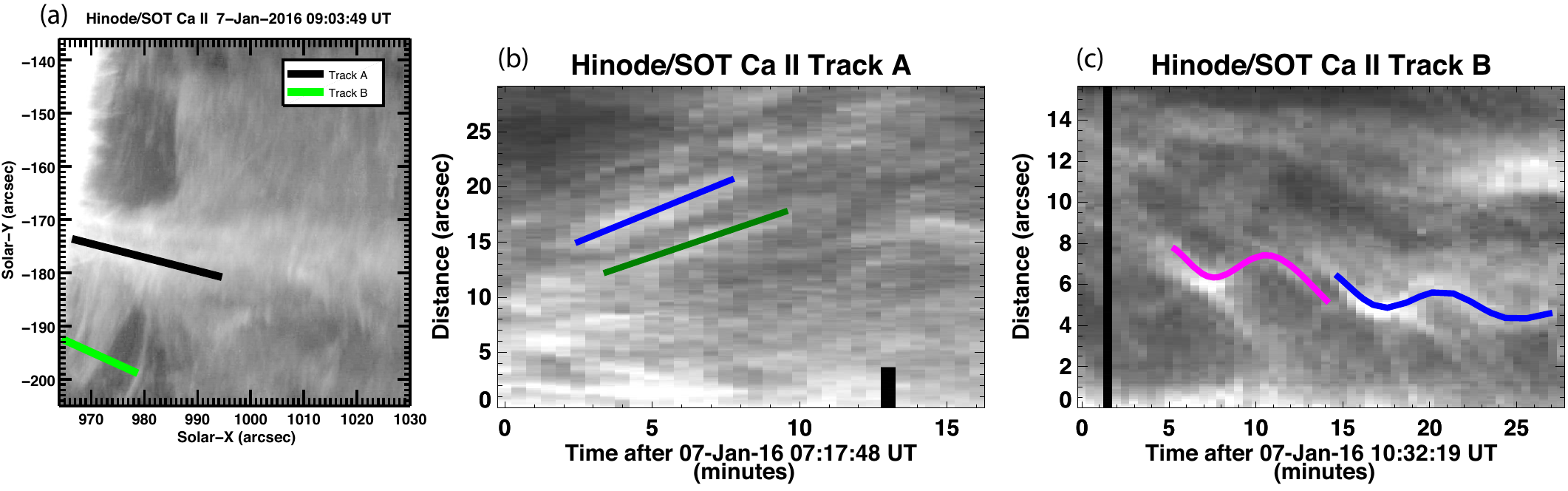}
}
\caption{(a) The location of the tracks perpendicular to the solar limb marked in Hinode/SOT image of the prominence on 7-Jan-2016 07:22:17 UT (Track A: black) and 10:42:49 UT (Track B: green). (b) Time distance diagram along Track A shown at left. Annotations show two motions along the slit with speeds in the plane of the sky from 10 to 13 km s$^{-1}$. (c) The time distance diagram along Track B annotated to show possible oscillations (solid pink and blue) with periods of $\sim$6-10 minutes.}
\label{CaIIvert:fig}
\end{figure}

We also measured broader features moving away from the solar limb as was done in \citet{KOT18} by inspecting the intensity vs. time along a particular tracks, as shown in Figure~\ref{CaIIvert:fig}. In Figure~\ref{CaIIvert:fig}a two tracks perpendicular to the solar limb  are marked (A, B). Track A is in the center of the pillar showing the propagating features, in agreement with previous studies \citep{OKKS15,KOT18}. The resulting time-distance diagram shown in Figure~\ref{CaIIvert:fig}b is for Track A, showing  features (green lines) with speeds from 10 to 13 km s$^{-1}$ in the plane of the sky, consistent with measurements discussed in \citet{KOT18} that can be identified as nonlinear fast magnetosonic waves \citep[see, also][]{OKKS15}.

Track B is outside the pillar, in the perpendicular direction to what appear to be propagating motions of cool prominence material. The time-distance plot (Figure~\ref{CaIIvert:fig}c) identifies a likely standing kink-mode  oscillations (solid lines) with periods of $\sim$6-10 minutes. For comparison, kink-mode oscillations of prominence threads were detected in several other observations in H$\alpha$ with periods  1.9-5.4 min using the Swedish 1m Solar Telescope in La Palma \citep{Lin09}, using Hinode/SOT with periods 3.5-8.75 min \citep{Nin09}, and in Ca~II with periods 2.9-4.2 min \citep{Oka07}  (see the review by \citet{Arr18}). In order to apply coronal seismology to the present observation of the kink oscillations we need to estimate the length of the oscillating thread, difficult primarily due to the projection and line of sight integration effects, as well as the small field of view of Hinode/SOT.

\section{Numerical Model and Boundary Conditions} \label{model:sec}

\begin{figure}[h]
\centerline{
\includegraphics[width=3.5in]{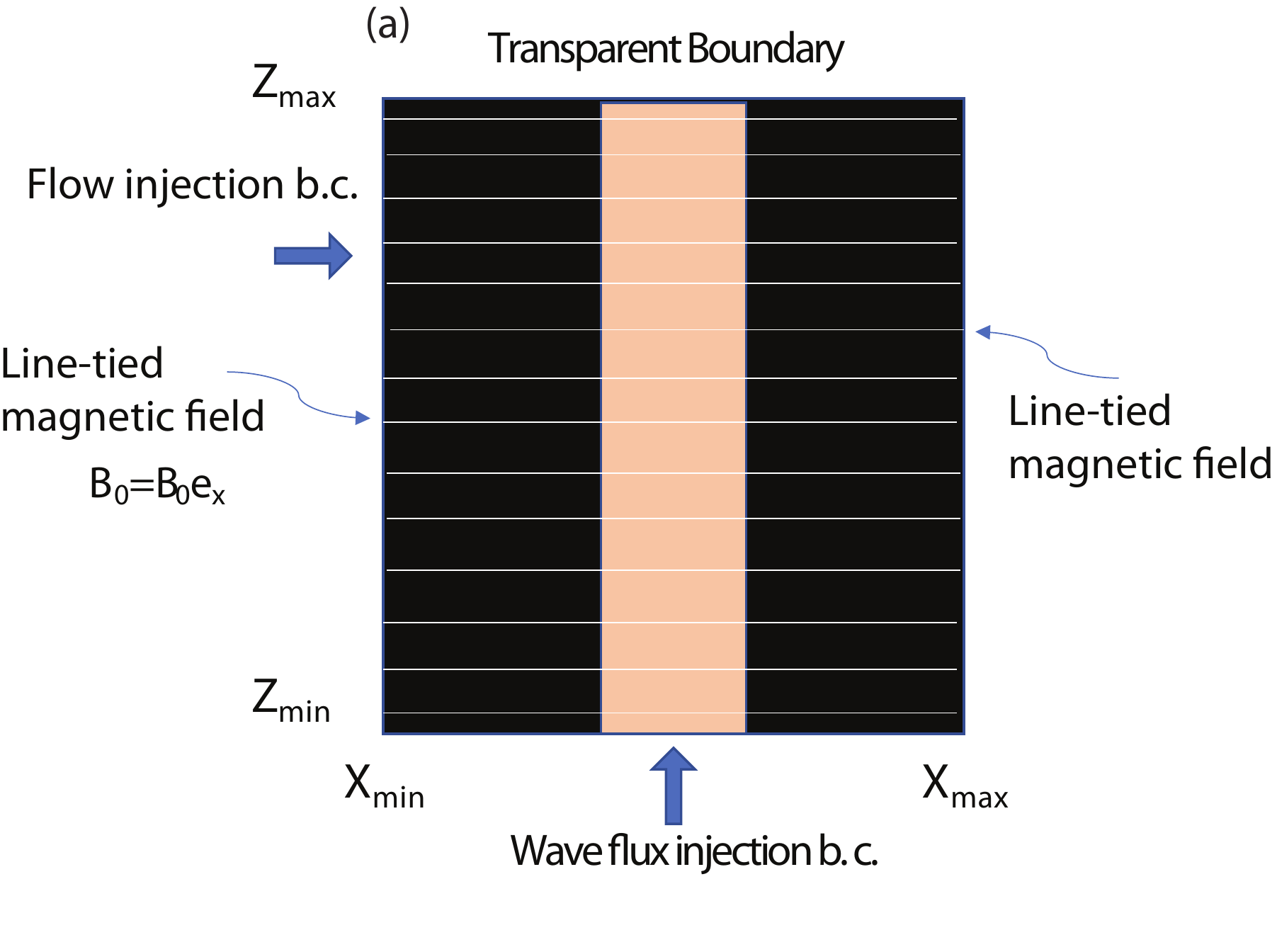}
\includegraphics[width=3.5in,height=2.75in]{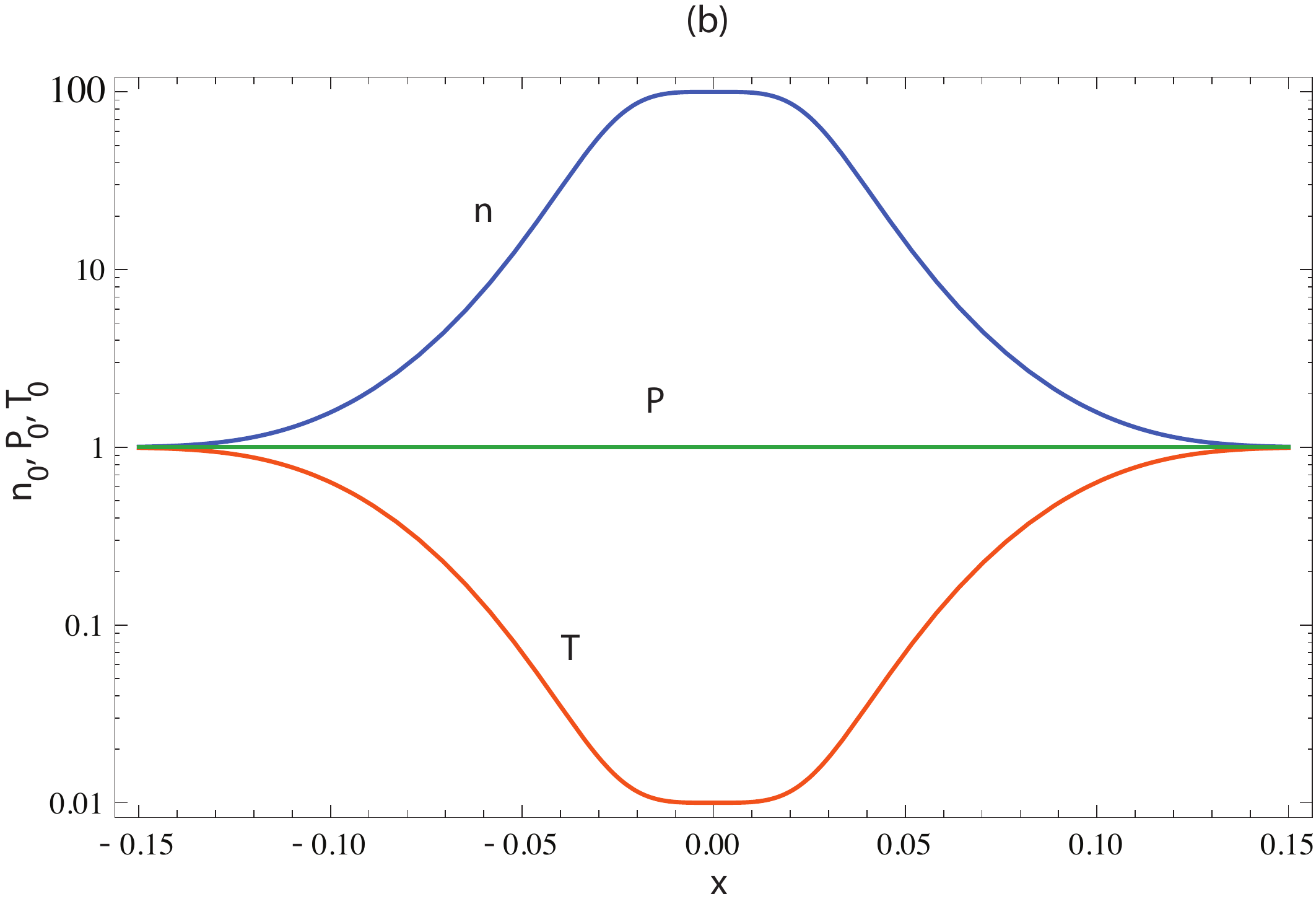}}
\caption{(a) The schematic diagram of the prominence foot model showing the high-density cool prominence material in the center, surrounded by coronal material, with horizontal magnetic field (white lines), and the various model boundary conditions (b.c.'s) of waves and localized flows injections. (b) The initial density, $n_0$ (blue), temperature, $T_0$ (red), and pressure $p_0$ (green) along a magnetic field line.}
\label{n0T0:fig}
\end{figure}

While we realize that the overall prominence structure is highly complex and dynamic, we employ a reductionist approach by using a simplified model that captures the physical properties of the fast magnetosonic waves and flows in narrow threads in the observed prominence foot. We use the 2.5D MHD model described in \citet{OKKS15} to model the fast magnetosonic waves and flows in a prominence foot using the initial state described below. Here, we repeat some basic properties of the  \citet{OKKS15} model for convenience. We apply the simplified model of \citet{JD92} to model the initial state of the prominence foot. The background magnetic field is horizontal with $\mbox{\bf B}=B_0\hat{x}$ (see, Figure~\ref{n0T0:fig}), and the background thermal pressure is in equilibrium (pressure balances), achieved by setting the following forms of temperature and density:  
\begin{eqnarray}
&&T(x,z)=T_{max}-(T_{max}-T_{min})e^{-[(x-x_0)/w]^{2p}},\\
&&n(x,z)=p_0/T(x,z),
\label{init:eq}
\end{eqnarray} where $T_{max}$ corresponds to the coronal temperature, and $T_{min}$ is the temperature of the prominence material inside the foot, $w$ is the half-width of the foot. In the present study we use $w=0.1$ normalized in units of $L_0$ (see below), $T_{min}/T_{max}=0.01$, and $p=2$ defines the sharpness of the profile.  Thus, the fast magnetosonic speed for perpendicular propagation $V_f=(V_A^2+C_s^2)^{1/2}$, where $V_A$ is the Alfv\'{e}n speed and $C_s$ is the sound speed, is lowest in the center of the foot and increases to the maximal value in the corona.  Since the magnetic field is strong (i.e., the magnetized plasma is low-$\beta$) and perpendicular to the direction of gravity, there is no initial gravitational stratification of the density in the model. The initial near-equilibrium state is show in Figure~\ref{n0T0:fig}b. However, during the evolution of the MHD model small gravitational stratification forms self-consistently, introducing height-dependence of the modeled quantities.

We solve the 2.5D resistive MHD equations with gravity in Cartesian geometry (with $y$-direction as the ignorable coordinate). The effects of the small resistivity are negligible on the timescale of the modeling runs. The following parameters are used: background magnetic field magnitude $B_0 = 5$ G (consistent with our estimate from observations in Section~\ref{obs:sec}), background coronal density $n_0=10^9$  cm$^{-3}$ coronal temperature $T_0 = 10^6$ K with corresponding  sound speed $C_s = 166$ km s$^{-1}$, and Alfv\'{e}n speed $V_A = 345$ km s$^{-1}$, Alfv\'{e}n time $\tau_A = 3.38$ min with distance normalization $L_0=70$ mM, and the thermal-to-magnetic pressure ratio $\beta=8 \pi p_0/B_0^2=0.2776$, where $p_0=k_B n_0T_0$, and $k_B$ is Boltzmann's constant, and the Lundquist number $S=10^4-10^5$.  We also use {\em B$_0$}=10 G in two runs (see, Table~1). The solutions are obtained in two spatial dimensions of the three components of the velocity and the magnetic field utilizing the fourth-order Runge–Kutta method in time and fourth-order spatial differencing on up to  $512^2$ grid points with open boundaries conditions at the top of the domain, time depended boundary conditions at the lower boundaries, and line-tied/inflow boundary conditions at the $x$-boundaries of the domain. A numerical convergence test was performed with the 2.5D MHD model to establish that the resolution is sufficient to resolve the modeled dynamical features. This was done by comparing runs with increasing grid resolutions until the results become independent of further resolution increase.

As in \citet{OKKS15}, fast magnetosonic waves are launched into the prominence foot at the lower boundary via time-dependent boundary condition on the $V_z$ component of the velocity. Motivated by the detected variability in the amplitude of  the waves, evident in Hinode/SOT animations, in time-distance, and in temporal plots that cut through the prominence pillar \citep[see, Figure 5 in][]{KOT18}, we apply a double frequency source, with the high $\omega_1$ frequency component  corresponding to the wave period of order several minutes, and a longer duration envelope component with an order of magnitude longer period. Thus, the corresponding velocity at the footpoint (i.e., the time dependent boundary condition) has the form
\begin{eqnarray}
V_z=\frac{V_{z,0}}{2}[{\rm cos}(\omega_1 t)+1] {\rm sin}(\omega_2t) e^{-[(x-x_0)/s_0]^4},
\label{vz0:eq}
\end{eqnarray} where we use $\omega_1=9\omega_2$. The typical velocity amplitude of the source is $V_{z,0}=0.015V_A$ corresponding to $\sim 10$ km s$^{-1}$, where $V_A$ is the local Alfv\'{e}n speed. The wave injection source is localized in the center of the pillar at $x=x_0$ with an approximate half width of $s_0=0.032$, falling off with the faster than exponential decay in the $x$-direction away from the center of the pillar.  Motivated by observations of localized flow in prominences we extend the previous 2.5D MHD model  \citet{OKKS15} by introducing an inflow boundary conditions at $t\ge 0$ at the $x$-direction (lateral along the background magnetic field) to inject field-aligned flow into the prominence foot region. The narrow Gaussian inflow at the boundary is of the following form
\begin{eqnarray}
V_x(x=x_{min},z,t)= V_{f0}e^{-[(z-z_k)/w_{in}]^2},
\label{vx0:eq}
\end{eqnarray} where $V_{f0}$ is the magnitude of the flow injected flow, the injection boundary is at $x_{min}$,  the normalized width $w_{in}=0.01$, the locations at $x_{min}$ boundary are at $z_k$, where, for example, at $k=1$, $z_{1}=h/2$, centered around the mid-height of the model prominence foot pillar. Additional  parameters of the model runs for each case considered here are given in Table~\ref{param:tab}.  Case 1 is modeled without the effects of gravity and with $B_0=5$ G. Cases 2 and 3 include the effects of gravity, and use $B_0=10$ G. In Case 3 counter-propagating flows are injected in narrow localized regions at the boundaries near the mid-height of the domain. Two overlaying flows are introduce on the left boundary towards the pillar, and a single flow thread is introduced on the right boundary in the opposite $x$-direction.

\begin{table}
\caption{The main parameters of the numerical model runs. The `$^*$' in Case 3 indicates counter-propagating flows injected from the left and right boundaries of the domain.}
\centering
\hspace{1cm}\begin{tabular}{cllccc}
\hline
Case \# & $V_{z,0}\ [V_A]$ & $\omega_1$ $[ Rad\tau_A^{-1}]$ & $V_{f0}\ [V_A]$  &  B [G] &   Gravity  \\ \hline
        1  &  0.015            &  10.61        &  0.0 &  5 &  No \\
        2 &  0.02               &  12.56      & 0.015      &    10     &     Yes  \\
       3 &    0.01            & 10.61       & $\pm$0.010$^*\ $          &   10  &   Yes   \\
      4 &     0.015              & 10.61        & 0.0        &  5       &  Yes\\
\hline
\hline
\end{tabular}
\\ 
\label{param:tab}
\end{table}

\section{Numerical Results} \label{num:sec}
\begin{figure}[h]
\centerline{
\includegraphics[width=\linewidth]{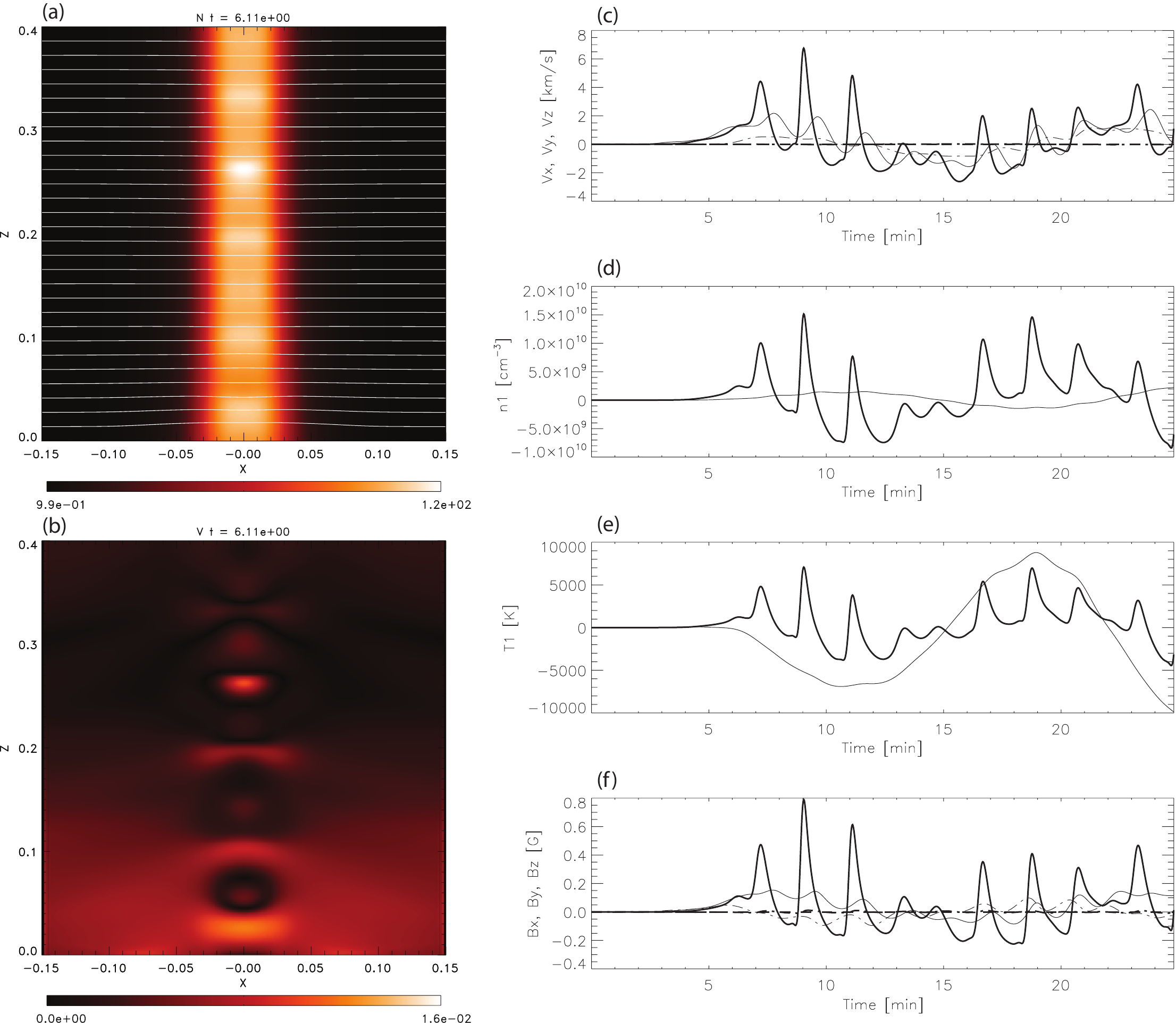}
}
\caption{The computation results obtained with the 2.5D MHD model for Case~1. (a) The magnetic field lines (white lines) and the density at $t=6.11\tau_A\approx 20.7$ min in units of $10^9$~cm$^{-3}$. (b) The corresponding velocity magnitude. An animation of this image is available online. The spatial dimensions are in units of $L_0=70$ mM. In the right column are the temporal evolutions of the variables at height $z=0.2$, the thick solid/dashed lines are at $x=0$ and the thin solid/dashed lines are at $x=0.05$.  (c) The velocity perturbation $V_z$ (solid), and the negligible $V_y$ (dashes), and $V_x$ components (dot-dashes). (d) The perturbed density, $n_1$. (e) The perturbed temperature $T_1$ in Kelvin.  Note, that at $x=0$ the values $T_1\times100$ are plotted in order to fit on the same scale with the values of $T_1$ at $x=0.05$. (f) The perturbed magnetic field component $B_x$ (solid), with small perturbations in $B_y$ (dashes) and $B_z$ (dot-dashes). The initial time-lag for the wave to reach the measurement height is evident. }
\label{f:Bnxz}
\end{figure}

The numerical solutions of the 2.5D equations with fast magnetosonic waves excited at the foot of the prominence pillar by time-dependent upward velocity, $V_{z0}(x,t)$ boundary condition, Equation~\ref{vz0:eq}, are shown in Figure~\ref{f:Bnxz} for Case~1. The spatial structure of the modeled pillar density and velocity is shown in Figure~\ref{f:Bnxz}a. The high density, cold pillar material is modeled in the central part of the region, and the values of the density are shown in the  color intensity scale. The nearly horizontal lines indicate the magnetic field lines, slightly deflected in the high density region due to wave pressure. The effect of the driven waves is seen as traveling density enhancements (see the enclosed animation), shown at $t=6.11\tau_A\approx 20.7$ min (with the present time normalization) and the corresponding propagating velocity fluctuations, in qualitative agreement with \textsl{Hinode}/SOT Ca~II observations. The effects of nonlinearity are evident in the steepening of the wave-fronts in agreement with the observed signatures of wave structures. The overall wave pressure is directed upwards along the pillar's foot that acts as a waveguide, and the waves  escape through the upper boundary of the modeled pillar, with open boundary conditions at the top. In the lower footpoint region the wave injection produces localized kink-like modes. A small fraction of the waves leaks outside the waveguide and propagate in the coronal region (low density, high-temperature region compared to the prominence pillar region).  We have repeated the run with the same parameters, but with gravity (Case~4), and found similar evolution of the waves. The main difference in this case arises from the gradual gravitational downward acceleration of the cool and heavy prominence material, and corresponding development of small curvature in the magnetic field lines for the case with $B_0=5$ G compared to the case without gravity. For the case with stronger ($B_0=10$ G, see below), the higher magnetic pressure diminishes further the effects of gravity perpendicular to the magnetic field.

\begin{figure}[h]
\centerline{
\includegraphics[width=0.58\linewidth]{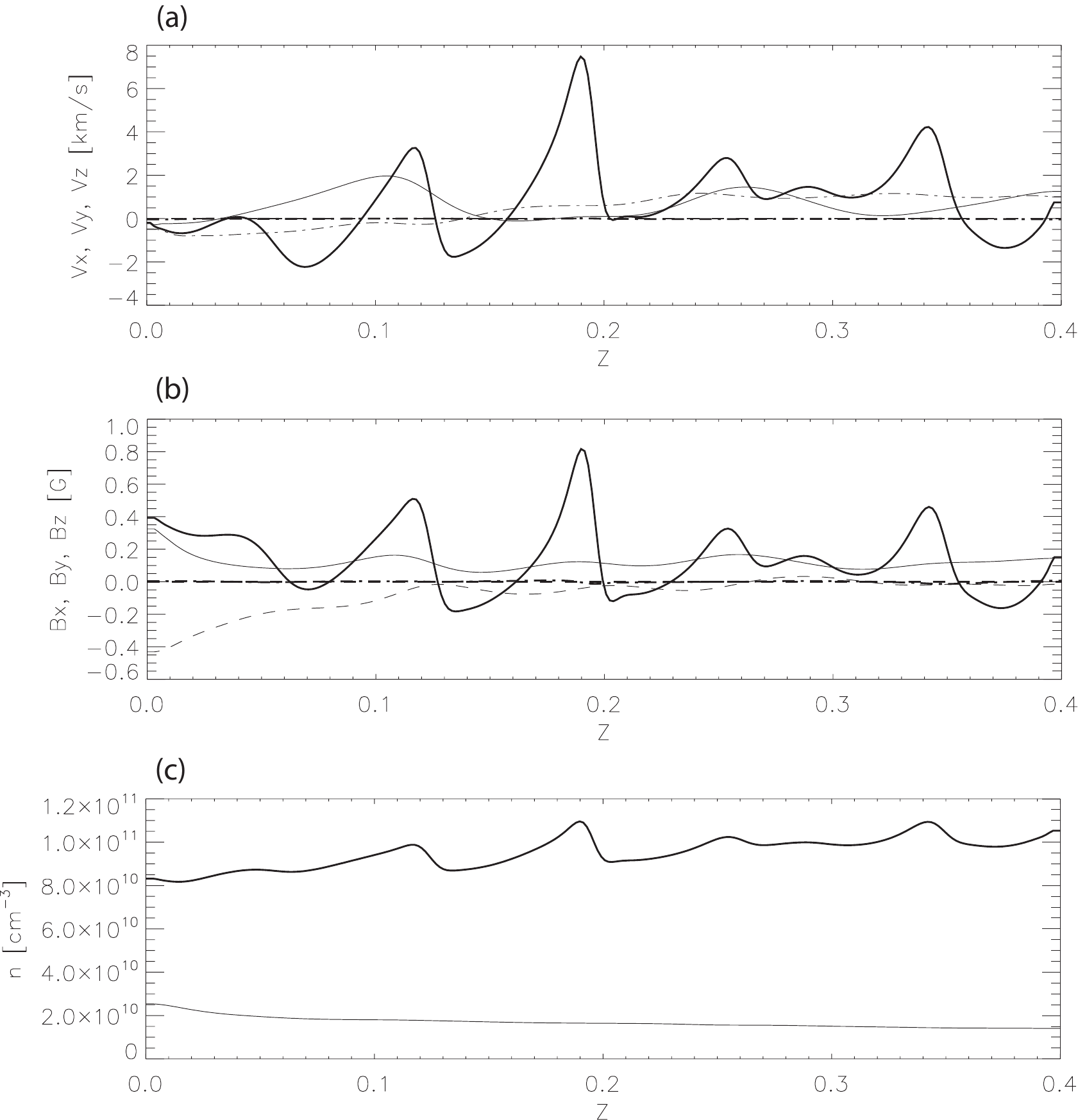}
}
\caption{ The cut at $x=0$ (thick solid/dashed lines) and at $x=0.05$ (thin solid/dashed lines) for (a) the velocity ($V_z$ (solid), $V_y$ (dashes) and $V_x$ (dot-dashes))components, (b) the perturbed magnetic field components ($B_x$ (solid), $B_y$ (dashes) and $B_z$ (dot-dashes)), and (c) total density $n$ for Case~1 shown in Figure~\ref{f:Bnxz}.}
\label{f:vbz}
\end{figure}

In Figure~\ref{f:Bnxz}c-f the temporal evolution of the variables at the center of the pillar ($x=0$) and near the  boundary of the pillar ($x=0.05$) at height $z=0.2L_0$ is shown. The initial time ($t\approx0.6\tau_A$) for the wave to reach the observed height is evident. Figure~\ref{f:Bnxz}c shows the temporal evolution of the velocity components. The $V_z$ component associated with the driven fast magnetosonic waves is dominant and exhibits the beating oscillations consistent with observations. The nonlinearity is evident in the steepened wave front and the non-sinusoidal structure of the waves, even though the injected source has a sinusoidal time-dependence. The density perturbation with respect to the initial pillar density, $n_1$, the temperature perturbations, $T_1$, oscillations are in phase with the $V_z$ and magnetic perturbation $B_x$ oscillations. These phase relations are consistent with the properties of the propagating fast magnetosonic waves (that are compressional).  The phase speed of the waves, $V_f$, increases at the boundary of the pillar towards the coronal region, and the fast magnetosonic waves undergo reflection and refraction in these regions, i.e., the wavefronts change direction and speed as evident in Figures~\ref{f:Bnxz}-\ref{f:vbz}, as well as coupling to other wave modes. This is evident by comparing the temporal evolution in Figure~\ref{f:Bnxz}c-f, and $z$-dependence (wavelength) of the waves at  $x=0$ and at $x=0.05$ in Figure~\ref{f:vbz}, where the phase shift of the velocity and magnetic field components is apparent, while the perturbed density and temperature are dominated by the long-wavelength perturbation at $x=0.05$ outside the pillar. 

\begin{figure}[h]

\plottwo{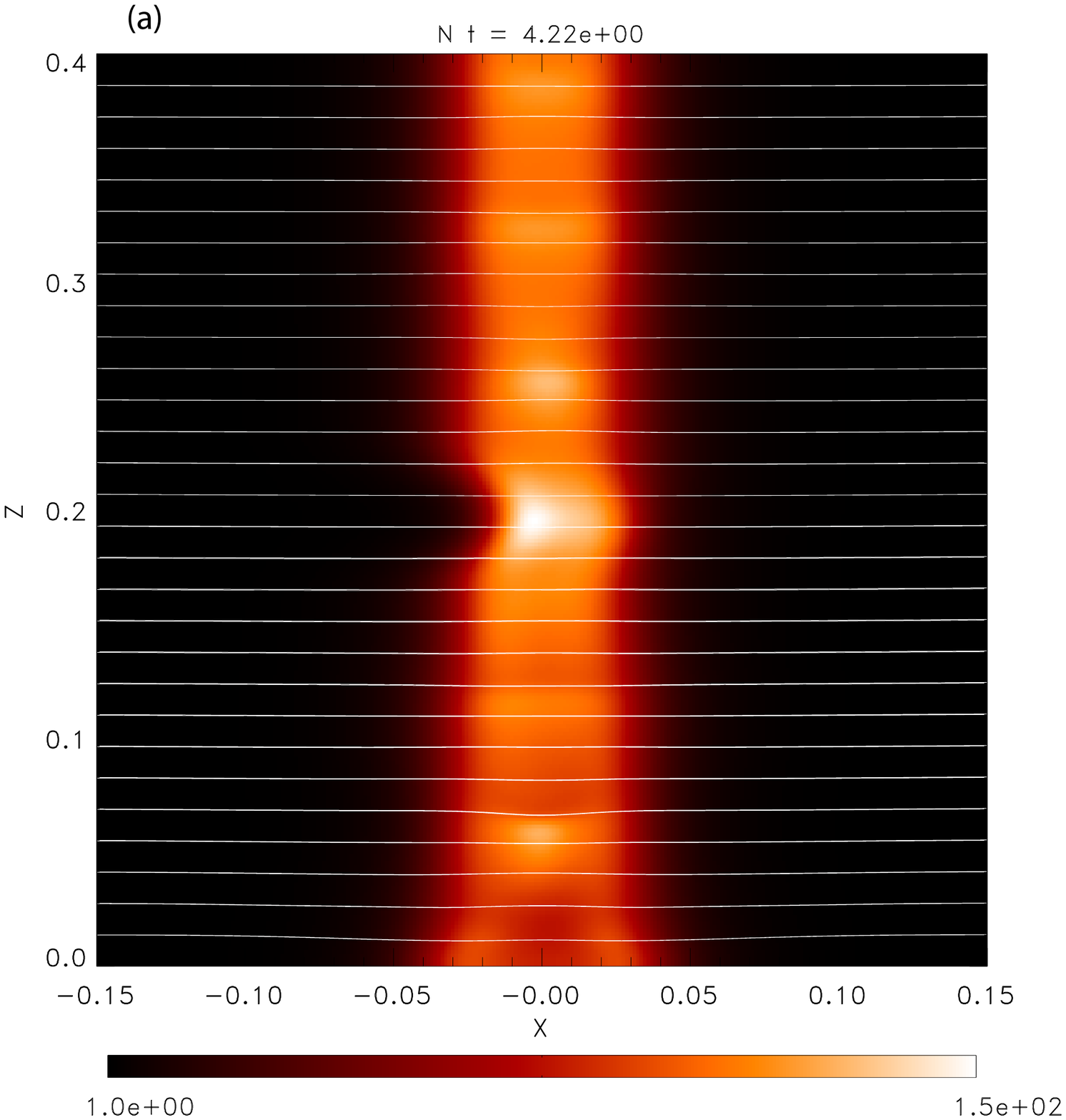}{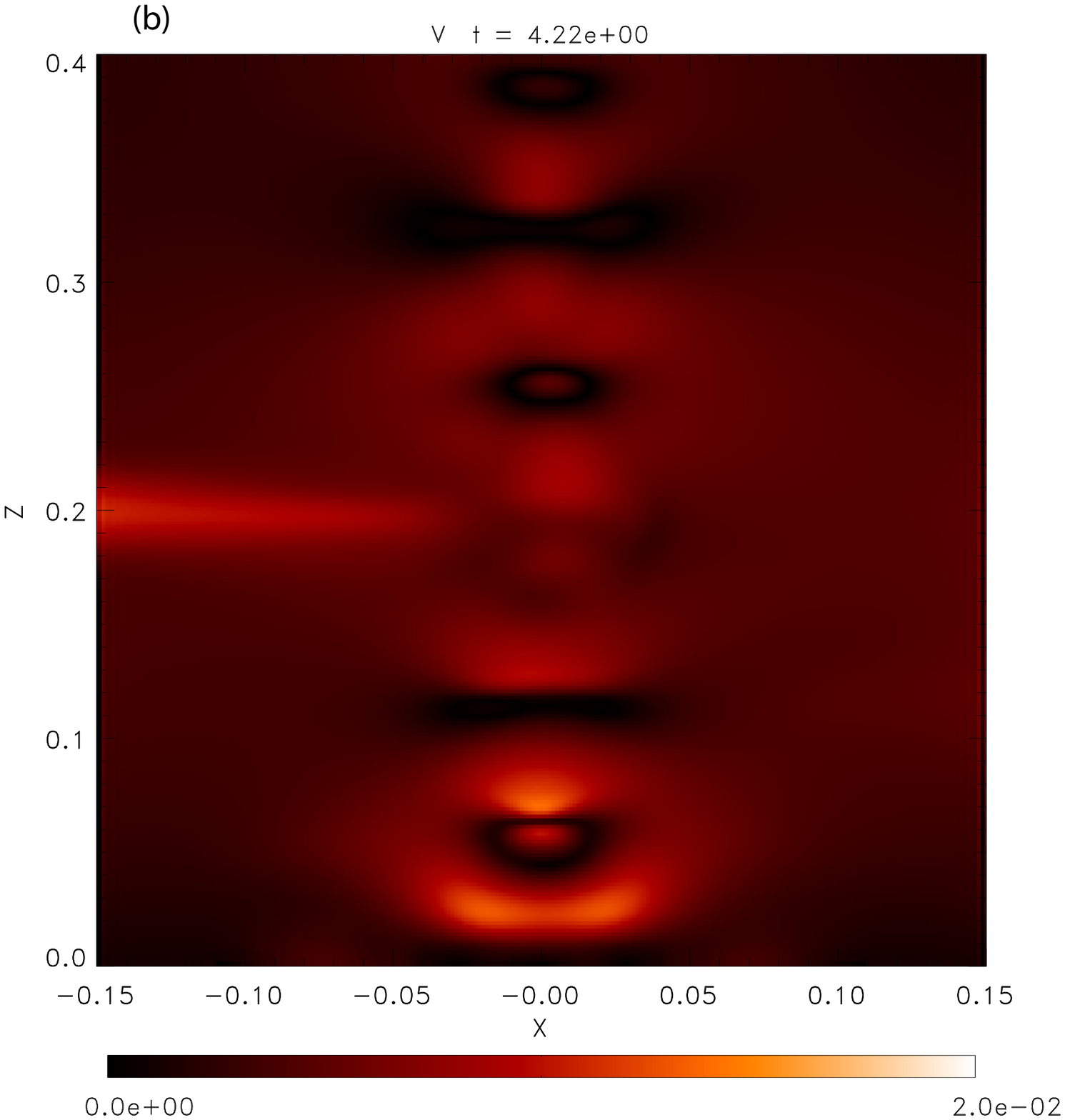}\vspace{-2cm}\plottwo{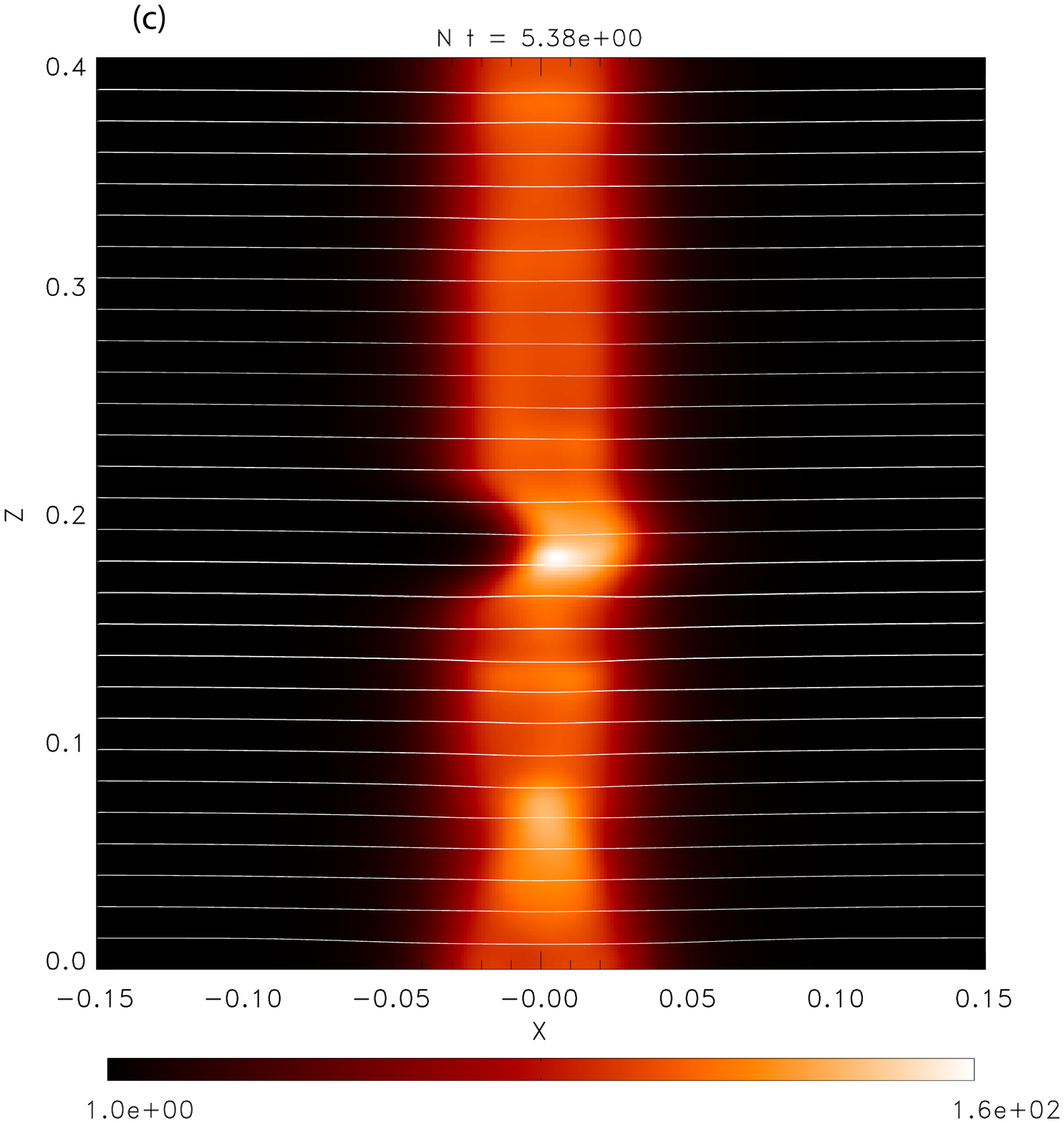}{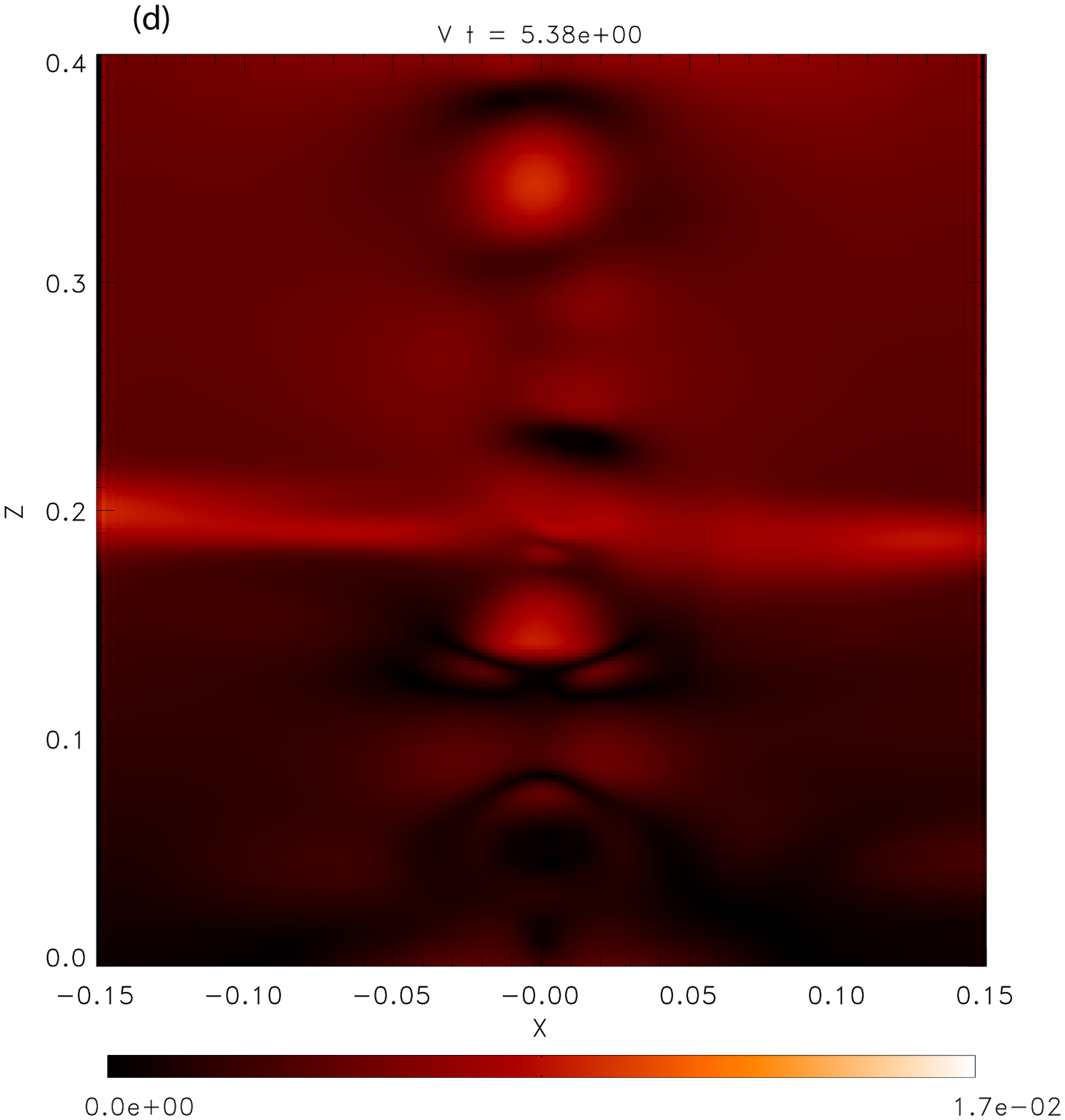}
\vspace{-1cm}\caption{The results of the 2.5D MHD model with waves and inflow along the magnetic field at the middle of the left boundary (Case2, normalized units). (a) the magnetic field lines (white) and the density at $t=4.22\tau_A$. (b)  the magnitude of the velocity $V$ in the $xy$ plane at $t=4.22\tau_A$. (c) same as (a) but at $t=5.38$. (d) same as (b) but at $t=5.38$. (An animation of this figure is available online article.) }
\label{f:Bnxz_v}
\end{figure}

In Figure~\ref{f:Bnxz_v} the results of the model for Case~2 at two times of the evolution are shown. In this case, the flow is injected at the left boundary of the model region with time-dependent velocity boundary condition given by Equation~\ref{vx0:eq}. In Figure~\ref{f:Bnxz_v}a, b the results are at $t=4.22\tau_A$, where the waves propagate to the upper boundary of the pillar, and the flow  has reached the pillar, causing the displacement of the high-density cold material, evident in Figure~\ref{f:Bnxz_v}a. The velocity magnitude, $V$, structure of the waves and the threaded flows  are shown in Figure~\ref{f:Bnxz_v}b. At later time,  $t=5.38\tau_A$ the flow has caused further mass deflection in the pillar (Figure~\ref{f:Bnxz_v}c) and the propagating material has reached the right boundary of the model (Figure~\ref{f:Bnxz_v}d). The simulation is run until the flow reflection from the opposite $x$-boundary reaches the pillar. It is evident that the velocity structure of the flow was affected by the interaction with the nonlinear waves and the changes the magnetic structure.

\begin{figure}[h]
\centerline{
\includegraphics[width=3.25in]{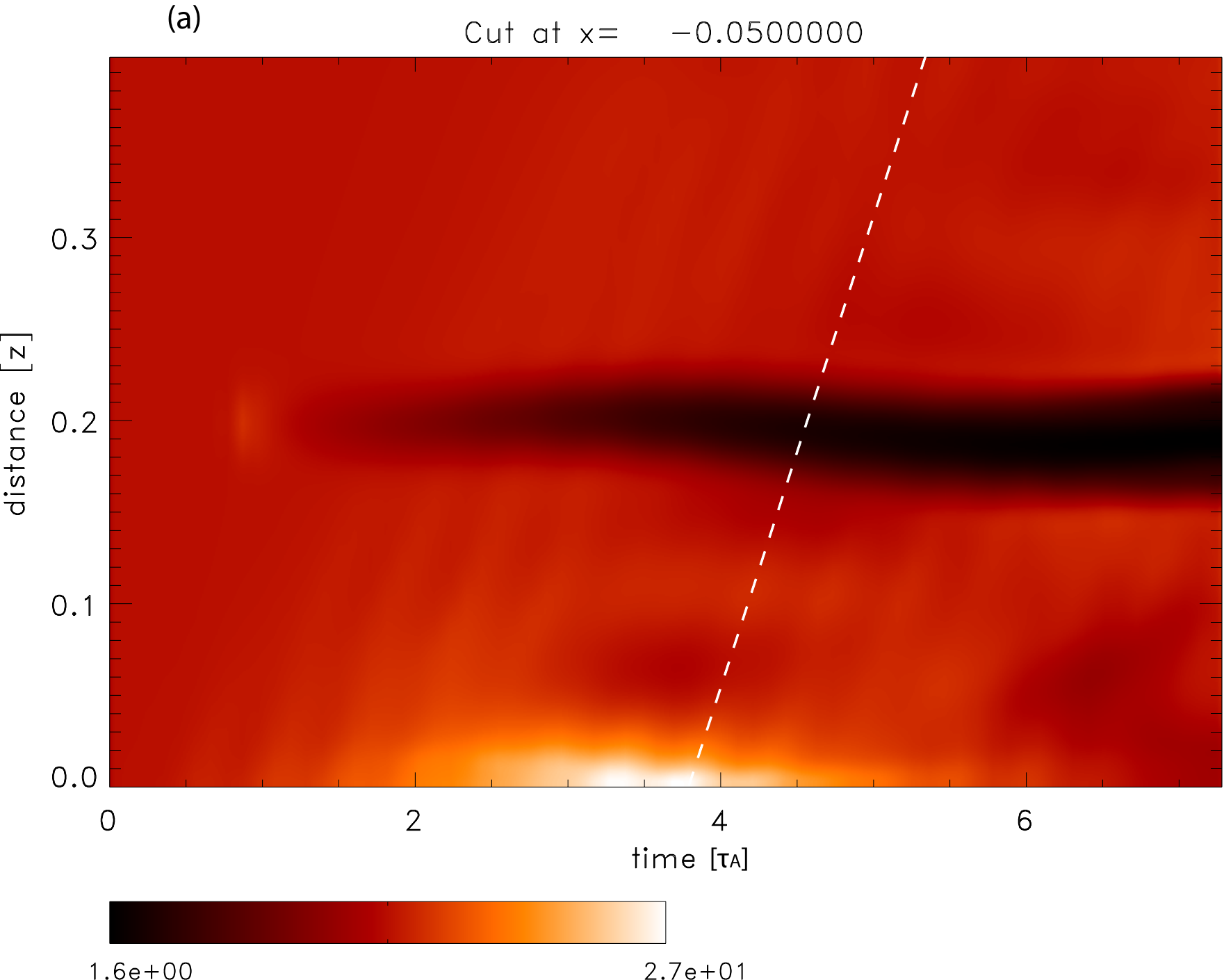}
\hspace{1cm}\includegraphics[width=3.25in]{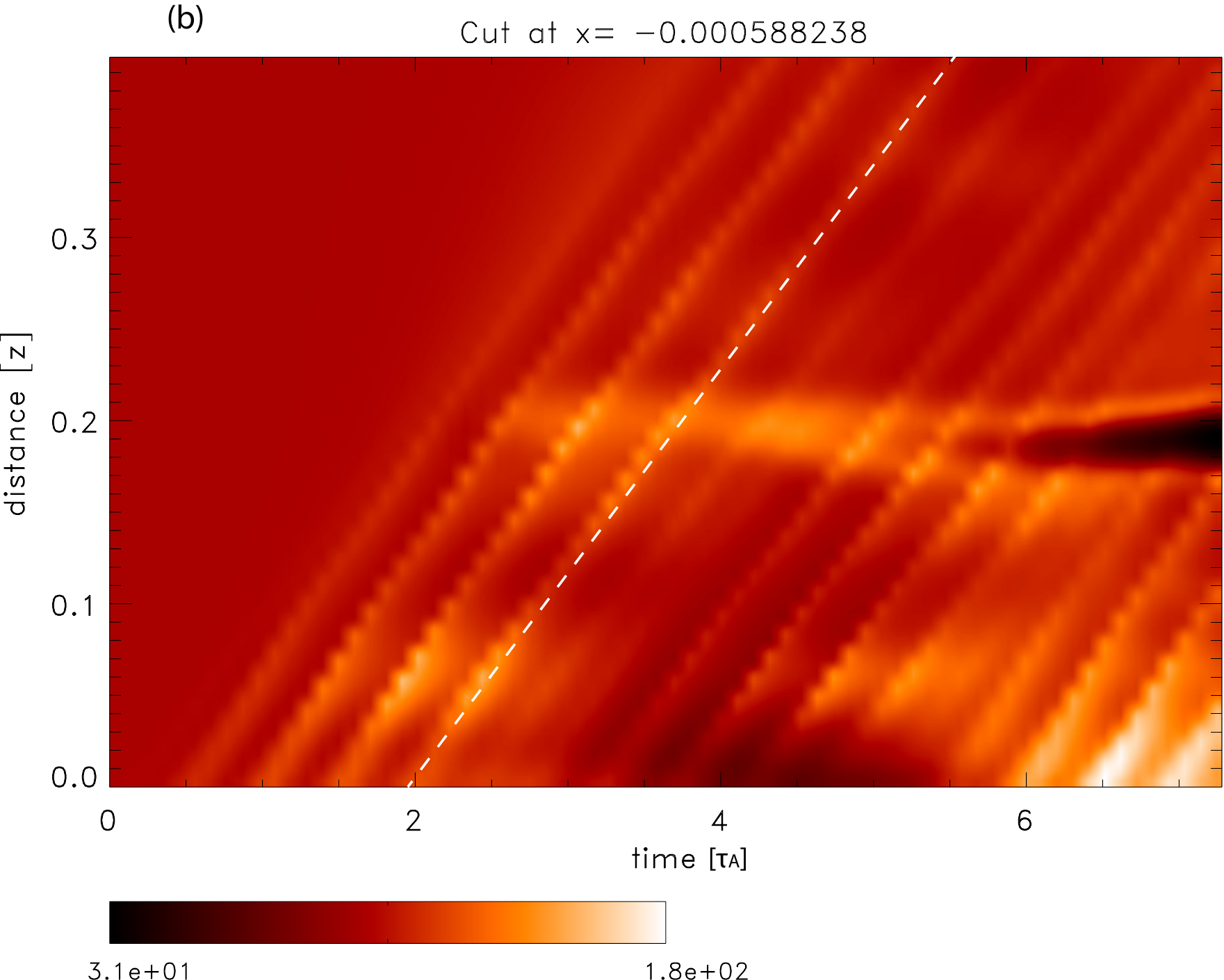}}
\caption{The time-distance plots (normalized units) of the density cuts from the 2.5D MHD model (a) at $x=0.05$ and (b) near the center of the pillar the fast magnetosonic speed is close to 0.1$V_A$, as expected for the high density low temperature region and is an order of magnitude smaller than in the coronal region. The speed in panel (a) of about  0.27$V_A$ is increased consistent with the density and temperature profiles shown in Figure~\ref{n0T0:fig}b. The dashed line show the velocity fit to the propagating fast magnetosonic waves features. }
\label{f:st}
\end{figure}

In Figure~\ref{f:st} the time-distance plots of the 2.5D MHD model results for the density are shown at two $x$-locations of the vertical cuts along the $z$-axis. In Figure~\ref{f:st}a the cut is to the left of the pillar, and the effects of the inflow velocity is evident in the density reduction due to the evacuated mass by the flow, and the transverse oscillations of the magnetic structure due to the time-variable injection of the waves by the time-dependent boundary condition  at $z-0$ (Equation~\ref{vz0:eq}). The small-scale variations due to the effects of the propagating waves are evident as  the alternating bright and dark slanted stripes with the slope corresponding to their propagation speed marked with the dashed line. 

The cut in the center of the pillar (near $x=0$) along the $z$-axis is shown in  Figure~\ref{f:st}b.  The nonlinear fast magnetosonic waves are well evident in this cut. Clearly, the phase velocity of these waves is diminished in the low-temperature, high-density pillar, compared to the high-temperature low-density region in  Figure~\ref{f:st}a with leaked waves. Near the center of the pillar the measured fast magnetosonic speed from the modeled waves is close to 0.1$V_A$ while at $x=0.05$ the measured speed increases to  0.27$V_A$, consistent with the density and temperature profiles shown in Figure~\ref{n0T0:fig}b that produce variation of the background fast magnetosonic speed along the magnetic field.  The effects of the inflow are seen by the density increase, followed by a decrease, and transverse oscillations due to the time-variable wave injection. The higher-frequency fast magnetosonic waves and lower frequency transverse oscillations model results are similar to the observational analysis results shown in Figure~\ref{CaIIvert:fig}.

\begin{figure}[ht]
\centerline{
\includegraphics[width=7in]{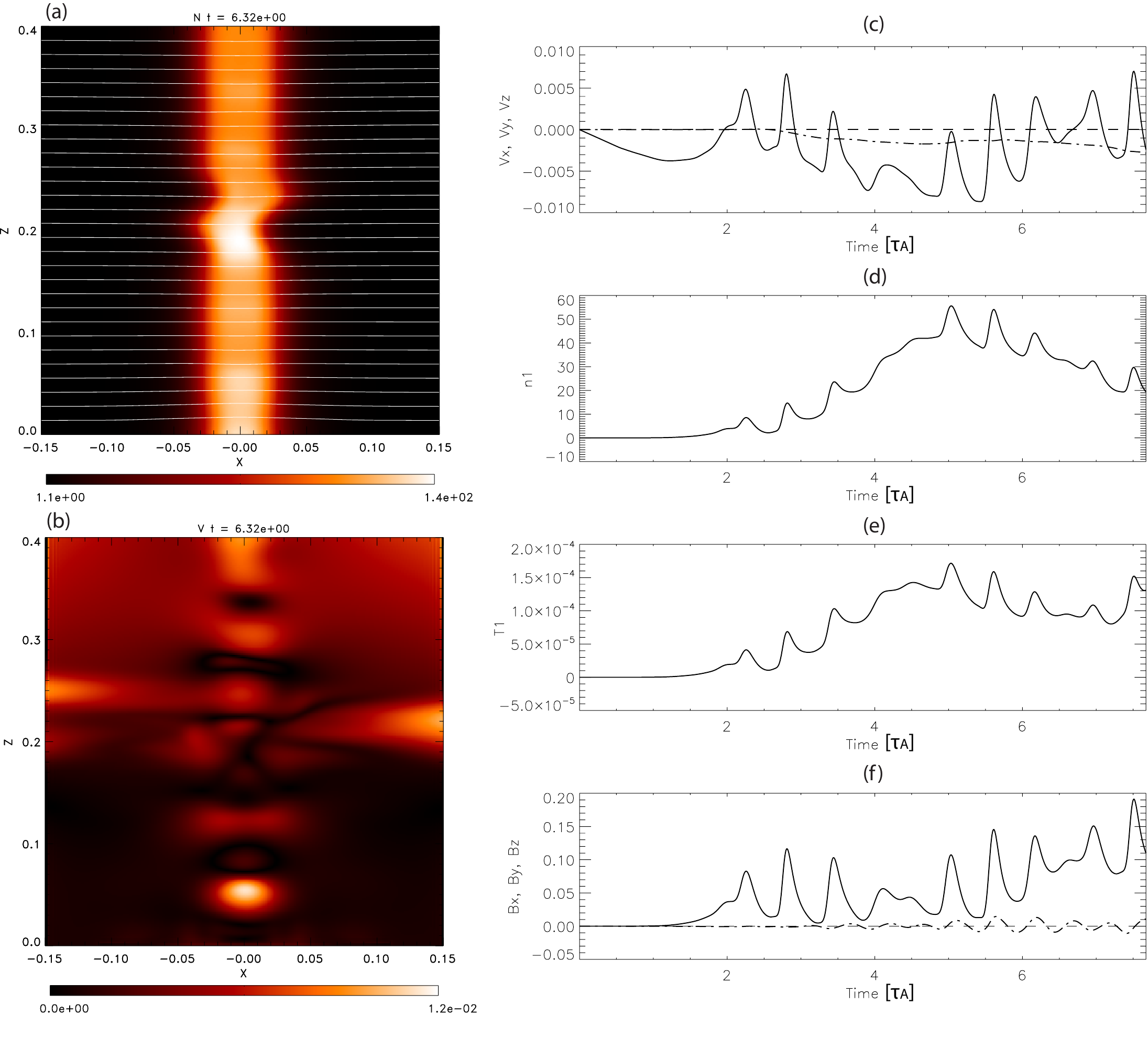}}
\caption{The effects of counter-propagating flows on the structure of the foot and on the fast magnetosonic waves. Same as Figure~\ref{f:Bnxz}, but for Case 3 (normalized units).}
\label{counter_v:fig}
\end{figure}

The generation of waves by counter-propagation flows and colliding flows observed by Hinode/SOT and IRIS SJI in highly energetic ($10^7-10^8$ erg cm$^{-2}$ s$^{-1}$) collision was studied and modeled by \citet{Ant18}. Here we model a two orders of magnitude lower energy event. Our study demonstrates the effects of background flows and density perturbation on the propagation of the nonlinear fast magnetosonic waves for injected as well as counter-propagating flows into the prominence foot (Figure~\ref{counter_v:fig}). For the counter propagating flows (Case 3) the flows are injected from two overlaying locations at the left boundary and from one location at the right boundary near the center of the height of the pillar. It is evident that the counter-propagating flows affect the structure of the foot by displacing the high-density cool material and introducing nonuniformity in the density structure of the prominence foot. It is evident that the fast magnetosonic waves are refracted in the nonuniform density region,  but still are trapped in the wave-guide structure and continue to propagate upward above the nonuniform region, in qualitative agreement with Hinode/SOT and IRIS SJI observations.

\section{Discussion and Conclusions} \label{dc:sec}
Recent high spatial and temporal resolution observations of a prominence foot with Hinode/ SOT Ca~II  and IRIS SJI Mg~II emissions show evidence of propagating features and oscillations. The structures are dynamic showing what appear to be motions both in the plane of the sky and along the line of sight. As discussed in \citet{KOT18}, the structures appear to be associated with shifted line components observed in the \ion{Mg}{2} spectral data. This is consistent with SOT H$\alpha$  data  showing red wing/ blue wing information that also suggests shifts associated with the bright moving features. These findings are consistent with flows and waves in a set of elongated field lines. The observations provide a diagnostic of the flows with velocities in the plane of the sky in the range of $\sim$16-47 km s$^{-1}$ in the plane of the sky in various parts of the prominence near the foot. In addition, the propagating nonlinear fast magnetosonic waves in the elongated prominence foot magnetic structures have velocities of $\sim$10-13 km s$^{-1}$. The transverse oscillations detected in the flowing prominence material threads have periods in the range of $\sim$6-10 min. Using a rough estimate $L= 50,000$ km for the length of the threads, based on their apparent structure, curvature and typical cool prominence material density of $10^{10}-10^{11}$ cm$^{-3}$, we use the expression for the kink speed \citep[e.g.,][]{NO01} 
\begin{eqnarray}
&&C_k=\frac{2L}{P}=\frac{B_0}{\sqrt{4\pi\rho_0}} \left(\frac{2}{1+\rho_e/\rho_0}\right)^{1/2},
\label{ck:eq}\end{eqnarray} where $P\approx6-10$ min is the oscillation period of the thread, $\rho_e$ is the coronal density and the quantities with subscript `0' are inside the thread. For simplicity, to estimate the magnetic field strength, $B_0$ we used the fact that the cool prominence material is much denser than the outside coronal material, $\rho_e\ll\rho_0$, and assumed the fundamental mode of oscillations of the threads. Thus, we find that $B_0\approx5-17$ G,  consistent with the typical magnetic field strength measured in prominence pillars \citep[e.g.][]{Sch13}.

Guided by the observed properties  of the waves and flows in the prominence foot we develop a simplified model of the wave propagation and oscillations guided by the prominence foot using the 2.5D MHD equations with the idealized initial state and time-dependent boundary conditions. The advantage of this approach in contrast to the complex observed prominence dynamics that includes a multitude of phenomena simultaneously, is the ability to model the oscillations processes and determine the linear and nonlinear waves modes, useful for coronal seismology. Using this model, we extend our previous studies of oscillating and  propagating features in the prominence foot  identified  as  guided nonlinear fast magnetosonic waves, and introduce the effects of flows along the direction of the magnetic field structures (or threads). The expanded, model demonstrates the effects of  flows on the cold high density prominence foot material, on the localized density structure in the foot, and on the propagation of the nonlinear fast magnetosonic waves. A single source of flows as well as counter-propagating flows were considered in the model. We found that the refracted waves in the flow-produced structures propagate in the prominence foot, guided along the high-density structure in agreement with Hinode/ SOT Ca~II  and IRIS SJI Mg~II emissions that show qualitatively similar evidence of flows and waves. 

In addition to propagating nonlinear fast magnetosonic waves,  the model reproduces transverse kink oscillations in the magnetic threads with flowing prominence material structures in agreement with our observations, and with previous observations of oscillating prominence threads in H$\alpha$ and Ca~II that were interpreted as kink modes. The modeling results help identify the physical nature of the observed oscillating and propagating features as nonlinear fast MHD and mass flows, and determine their effects on the structure of the prominence foot. Moreover, the model further validates the application of coronal seismology to the observed prominence structures.

\acknowledgments

LO acknowledges support by NASA Cooperative Agreement
NNG11PL10A to CUA. TAK acknowledges support by the NASA HGI program. Hinode is a Japanese mission developed and launched by ISAS/JAXA, with NAOJ as domestic partner and NASA and STFC (UK) as international partners. It is operated by these agencies in co-operation with ESA and NSC (Norway). IRIS is a NASA small explorer mission developed and operated by LMSAL with mission operations executed at NASA Ames Research center and major contributions to downlink communications funded by ESA and the Norwegian Space Centre. 

%

\vspace{5mm}
\facilities{Hinode(SOT), IRIS}

\end{document}